\documentclass[aps,pra,twocolumn,amsmath,amssymb,nofootinbib,superscriptaddress]{revtex4-1}

\usepackage{times}
\usepackage[pdftex]{graphicx}
\usepackage{dcolumn}
\usepackage{bm}
\usepackage{amsmath}
\usepackage{indentfirst}
\usepackage{float}
\usepackage[colorlinks]{hyperref}
\usepackage[dvipsnames]{xcolor}
\usepackage{xcolor}
\usepackage{subfigure}
\usepackage{algorithm}
\usepackage{algorithmicx}
\usepackage{algpseudocode}
\usepackage{amsmath}
\usepackage{verbatim}
\usepackage{makecell}
\usepackage[normalem]{ulem}

\newcommand{\loss}{{\rm loss}}

\newcommand{\ppt}{\hat{\Upsilon}}
\newcommand{\pptd}{\hat{\Upsilon}^{\dagger}}
\newcommand{\pt}{\Upsilon}
\newcommand{\trace}{{\rm tr}}

\newcommand{\evo}{\hat{\mathcal{U}}^{SE}}
\newcommand{\transfer}{E}

\newcommand{\mem}{\mathcal{C}_q}
\newcommand{\rhoss}{\rho_{st}}

\newcommand{\hnu}{Key Laboratory of Low-Dimensional Quantum Structures and Quantum Control of Ministry of Education, Department of Physics and Synergetic Innovation Center for Quantum Effects and Applications, Hunan Normal University, Changsha 410081, China
}

\begin{document}

\title{Memory Complexity of Quantum Processes}

\author{Chu Guo}
\email{guochu604b@gmail.com}
\affiliation{Henan Key Laboratory of Quantum Information and Cryptography, Zhengzhou,
Henan 450000, China}
\affiliation{\hnu}



\begin{abstract}
Generic open quantum dynamics can be described by two seemingly very distinct approaches: a top down approach by considering an (unknown) environment coupled to the system and affects the observed dynamics of the system; or a bottom up approach which tries to build an open quantum evolution model from the observed data. The process tensor framework describes all the possible observations one could possibly make on a quantum system, however it is computationally inefficient and not predictive. Here we define the purified process tensor which 1) allows efficient tomography as well as prediction for future process and 2) naturally defines a stationary quantum process as well as a quantitative and easy-to-evaluate definition of the memory complexity, or the degree of non-Markovianity, for it. As such it allows to uncover the minimal open quantum evolution model hidden in the observed data, completing the second approach for understanding open quantum dynamics. The intimate connection between quantum processes and classical stochastic processes is drawn in the end. 
\end{abstract}

\maketitle

\section{Introduction}

A quantum system is almost inevitably affected by some environment, in which case the dynamics has to be described in the context of an open quantum system~\cite{InesAlonso2017}. A powerful tool to study open quantum system is the Quantum Map, which is a linear and completely positive (CP) mapping from a quantum state at time $t_0$ to another quantum state a later time $t_1$~\cite{SudarshanRau1961,JordanSudarshan1961}. However, in cases where the initial states of the system plus environment are entangled, Quantum Maps are not properly defined since one can not arbitrarily prepare linearly dependent initial states of the system without altering the environment~\cite{Pechukas1994,Alicki1995}. Nevertheless, this situation can be circumvented if one considers the mapping from all possible experimental operations, defined as a generalized instrument~\cite{DaviesLewis1970}, to the output quantum state instead. The latter mapping is referred to as a superchannel which preserves both linearity and complete positivity~\cite{MilzModi2017}. 

Generalization of such a superchannel to multiple times naturally leads to the concept of the process tensor framework, defined as a mapping from multiple instruments at different times into an output quantum state~\cite{CostaShrapnel2016,PollockModi2018a}. Process tensor is the most general description of a multi-time quantum process, which is purely based on experimentally measurable data, in comparison to the open quantum evolution (OQE) description. It resolves at least two conceptual difficulties in open quantum theories: 1) it preserves both linearity and complete positivity in presence of initial system-environment entanglement and 2) it gives a clear and intuitive definition of Markovianity for quantum processes~\cite{PollockModi2018b}, based on which it is shown that a CP divisible quantum process may not be Markovian~\cite{MilzModi2019}. Process tensors with a few time steps (no more than $3$) have been reconstructed experimentally~\cite{WhiteModi2020,XiangGuo2021,GoswamiCosta2021}.

Given the current progresses on the process tensor framework, there are two important aspects which are not satisfactorily addressed yet: 1) a quantitative and easy-to-evaluate definition of non-Markovianity is in lack, the definition in Ref.~\cite{PollockModi2018b} does not lend itself to practical evaluation, which does not have a clear classical correspondence either; 2) it is not predictive. Even if one has reconstructed the process tensor till a time step $N$, it does not give much insight into the quantum process at later times. In addition, the process tensor is shown to be able to be efficiently represented as an matrix product density operator (MPDO) for finite-size environment~\cite{PollockModi2018a}, however, an efficient and deterministic tomography algorithm for a generic MPDO is still unknown to the best of our knowledge, without additional assumptions as made in Ref.~\cite{BaumgratzPlenio2013}. This is in contrast with the case of pure states that can be represented as Matrix Product States (MPSs), in the latter case polynomial tomography algorithms are known~\cite{CramerLiu2010,LanyonRoos2017}. To this end, we also note that in certain cases where the system-environment dynamics are explicitly known, the process tensor can be systematically computed as an Matrix Product Operator (MPO)~\cite{GuoPoletti2020,JorgensenPollock2019}, even if the environment may contain an infinite number of degrees of freedom (DOFs). An MPO based machine learning algorithm has also been proposed to reconstruct the process tensor based on only a polynomial number of experimental measurements, however the MPO ansatz does not preserve complete positivity and this algorithm is not deterministic, thus can not guarantee success in general~\cite{GuoPoletti2018,GuoPoletti2020}.

These issues are addressed in this work, under the assumption of a finite-size environment. First we define the purified process tensor (PPT), which is closely related to the process tensor but is a pure state instead of a density operator. We show that PPT is a sequentially generated multi-qubit state~\cite{SchonWolf2005} whose bond dimension equals to the environment size.
As a result, PPT has several advantages over the process tensor: 1) It allows efficient tomography with a slightly modified algorithm based on Ref.~\cite{CramerLiu2010}. Moreover, with the reconstructed PPT, one can immediately uncover the minimal OQE hidden in the quantum process. In case the OQE is time-independent, one could use it to predict any future process; 2) It is deeply connected to the $\epsilon$-machine for classical stochastic process~\cite{CrutchfieldYong1989,ShaliziCrutchfield2001}, based on which we could naturally define a \textit{stationary quantum process} as well as the \textit{memory complexity} of it. The exact values of the memory complexities for generic quantum processes are given.

\begin{figure}
\includegraphics[width=\columnwidth]{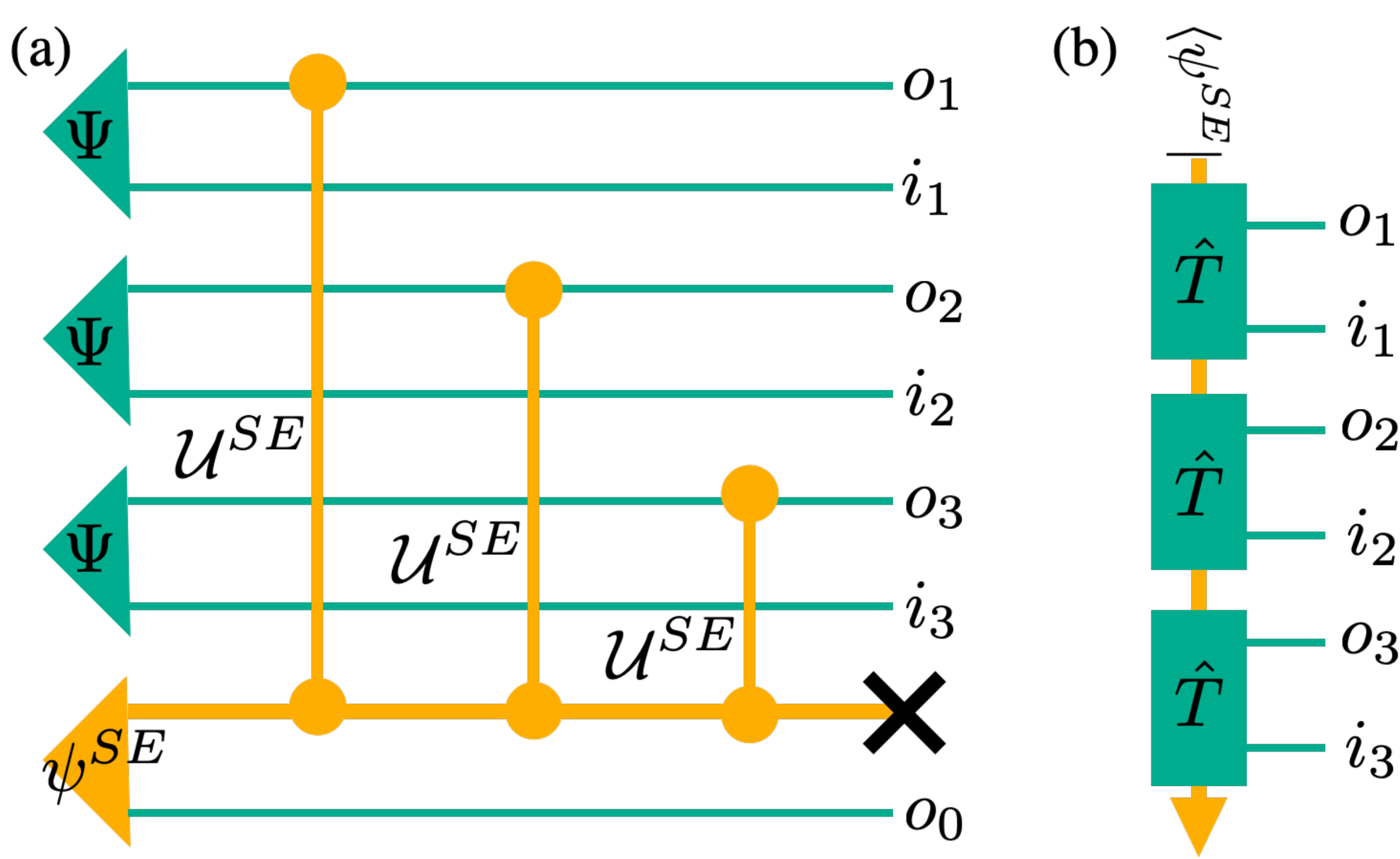}
\caption{(a) The quantum circuit which generates the Choi representation of the process tensor, with $\evo$ the evolutionary operator acting on the system and the environment. The purified process tensor is the full output state of the circuit, while the process tensor is obtained by tracing out the thick black leg corresponding to the environment. $\vert \Psi\rangle$ is the maximally entangled state and $\vert \psi^{SE}\rangle$ is the system-environment initial state. (b) Purified process tensor as a sequentially generated multi-qubit state. $\hat{T}$ is related to $\evo$ by Eq.(\ref{eq:Tmat}).
}
\label{fig:fig1}
\end{figure}

\section{Purified process tensor}
We consider a system (denoted as $S$) which undergoes a discrete quantum process. A most generic description for such a process is the OQE, in which the system is coupled to some environment (denoted as $E$), and the system plus environment undergoes unitary dynamics between two time steps $n$ and $n+1$ with the unitary evolutionary operator $\evo_{n}$. The case that $\evo_{n}$ is independent of $n$ is referred to as the time-independent quantum process. In the next we will omit the subscript $n$ for briefness whenever it does not affect the context. In general, $\evo$ can be written as $\evo \vert j, \alpha\rangle = \sum_{k, \beta} U^{j, \alpha}_{k, \beta} \vert k, \beta\rangle$, where $\vert j\rangle$, $\vert k\rangle$ are states for the system and $\vert \alpha\rangle$, $\vert \beta\rangle$ are states for the environment. $U^{j, \alpha}_{k, \beta}$ is a rank-$4$ tensor corresponding to $\evo$ in the specifically chosen basis. It is shown that for an $N$-step quantum process, the Choi representation of the process tensor, denoted as $\pt_{N:1}$, is a rank-$2N+1$ tensor that can be represented as an MPDO. Moreover, it can also be generated by evolving the quantum circuit as shown in Fig.~\ref{fig:fig1}(a)~\cite{PollockModi2018a}, where $i_n$ and $o_n$ denote the $n$-th input and output indices of the Choi state. Since we will only work with the Choi state of the process tensor and that the original process tensor and its Choi representation can be equivalently transformed into each other, we will not distinguish them and simply refer to the Choi state as the process tensor.

A key observation is that although the process tensor is a density operator, the quantum circuit in Fig.~\ref{fig:fig1}(a) is unitary, therefore the output state is a pure state if the environment is not traced out. We define this pure state as the purified process tensor (PPT). We show that PPT allows to address those issues mentioned in the introductory part, which are difficult or impossible to be addressed with the process tensor. First, the quantum circuit that generates PPT implies that PPT is a sequentially generated multi-qubit state of the form~\cite{SchonWolf2005}
\begin{align}\label{eq:ppt}
\ppt_{N:1} \vert \psi^{SE}\rangle =& \hat{T}_N \dots \hat{T}_2 \hat{T}_1 \vert \psi^{SE} \rangle,
\end{align}
except that the environment is not decoupled from the system in the end. Here $\vert \psi^{SE}\rangle$ is the initial state of the system plus environment and $\ppt_{N:1} \vert \psi^{SE}\rangle$ denotes the PPT which is a pure state that has $2N+1$ indices, with $2N$ `physical' indices corresponding to $\{o_1, i_1, \dots, o_N, i_N\}$ and one `auxiliary' index corresponding to the system plus environment. $\hat{T}_n$ is an isometry acting on the environment and the $i_n$, $o_n$ indices. The PPT is also shown in Fig.~\ref{fig:fig1}(b), which is naturally an MPS with bond dimension $D$ (assuming $D$ is the environment size). The explicit form of $\hat{T}_n$ can be seen from the construction of the quantum circuit, that is, assuming that $\hat{V}$ is the unitary operation which maps the state $\vert 00\rangle$ into the maximally entangled state $\vert \Psi\rangle = \sum_{j=1}^d \vert jj\rangle / \sqrt{d}$, then we have
\begin{align}\label{eq:Tmat}
\hat{T} \vert \alpha\rangle \vert 00\rangle =& \evo \hat{V} \vert \alpha\rangle \vert 00\rangle = \frac{1}{\sqrt{d}} \sum_{\beta, k, j} U^{j, \alpha}_{k, \beta} \vert \beta, k, j\rangle ,
\end{align}
where we have omitted the subscript $n$ for both $\hat{T}$ and $\evo$. The site matrix $B$ of the corresponding MPS representation of PPT can be read from Eqs.(\ref{eq:ppt}, \ref{eq:Tmat}) as $B_{\beta, \alpha}^{j, k} =  U^{j, \alpha}_{k, \beta} / \sqrt{d} $, which simply reorders the indices of the tensor $U$. It is often convenient to view $B_{\beta, \alpha}^{j, k}$ as a list of matrices acting on the environment DOFs, labeled by the physical indices $j,k$. It can be seen that PPT is invariant under any isometry $\hat{S}$ acting on the environment, that is, changing the basis of the environment or mapping them into another set of basis of a larger environment. Moreover, applying an additional unitary transformation onto the output environment state of PPT does not have any observable effects since it will be contracted out. The details of those properties as well as the isometry property of $\hat{T}$ are proven in the Supplementary. 

To this end we note that in our definition of PPT the initial state of the system, which corresponds to the leg labeled by $o_0$ in Fig.~\ref{fig:fig1}(a), is treated on the same footing as the environment, instead of as one of the physical indices such as $i_n$ and $o_n$. As a result the later process starting from the first time step feels an effective environment with size $D' = d\times D$ ($d$ is the system size). Therefore $\hat{T}$ in Eq.(\ref{eq:ppt}) should be understood as $\hat{T} \otimes \hat{I}^S$, with $\hat{I}^S$ the identity operator acting on the index $o_0$. If the $\vert \psi^{SE}\rangle$ is not entangled, that is, $\vert \psi^{SE}\rangle = \vert \psi^S\rangle \otimes \vert \psi^E\rangle$ with $\vert \psi^S\rangle$ and $\vert \psi^E\rangle$ initial states for the system and environment respectively, then Eq.(\ref{eq:ppt}) can be simplified as $\ppt_{N:1} \vert \psi^{S}\rangle \otimes \vert \psi^E \rangle  = \left(\ppt_{N:1} \vert \psi^E \rangle\right) \otimes \vert \psi^{S}\rangle$, thus $\vert \psi^S\rangle$ can be split off from the effective environment, and the resulting PPT will only have bond dimension $D$. Such a treatment allows a more uniform MPS representation since $o_0$ is not acted on by $\evo$, it also more transparently reveals the fact that the initial system-environment entanglement will in general increase the non-Markovianity (the exact meaning of which will become clear later) of the quantum process since the size of the effective environment is larger. In the following the environment corresponding to an entangled initial state will always mean the effective one and $D$ should be understood as $D'$ accordingly, unless particularly specified. 
Compared to the process tensor, the probabilities of future process conditioned on the measurement outcome of the initial state is lost. Nevertheless, this piece of information can also be restored with PPT (See Supplementary). For separable $\vert \psi^{SE}\rangle$ the future process is independent of any measurements performed on the initial state of the system, then the MPDO form of the process tensor is related to our definition of PPT by $\pt_{N:1} = \trace_E( \ppt_{N:1} \vert\psi^E\rangle\langle \psi^E\vert \pptd_{N:1} )$.


\section{Efficient purified process tensor tomography}
\begin{figure}
\includegraphics[width=\columnwidth]{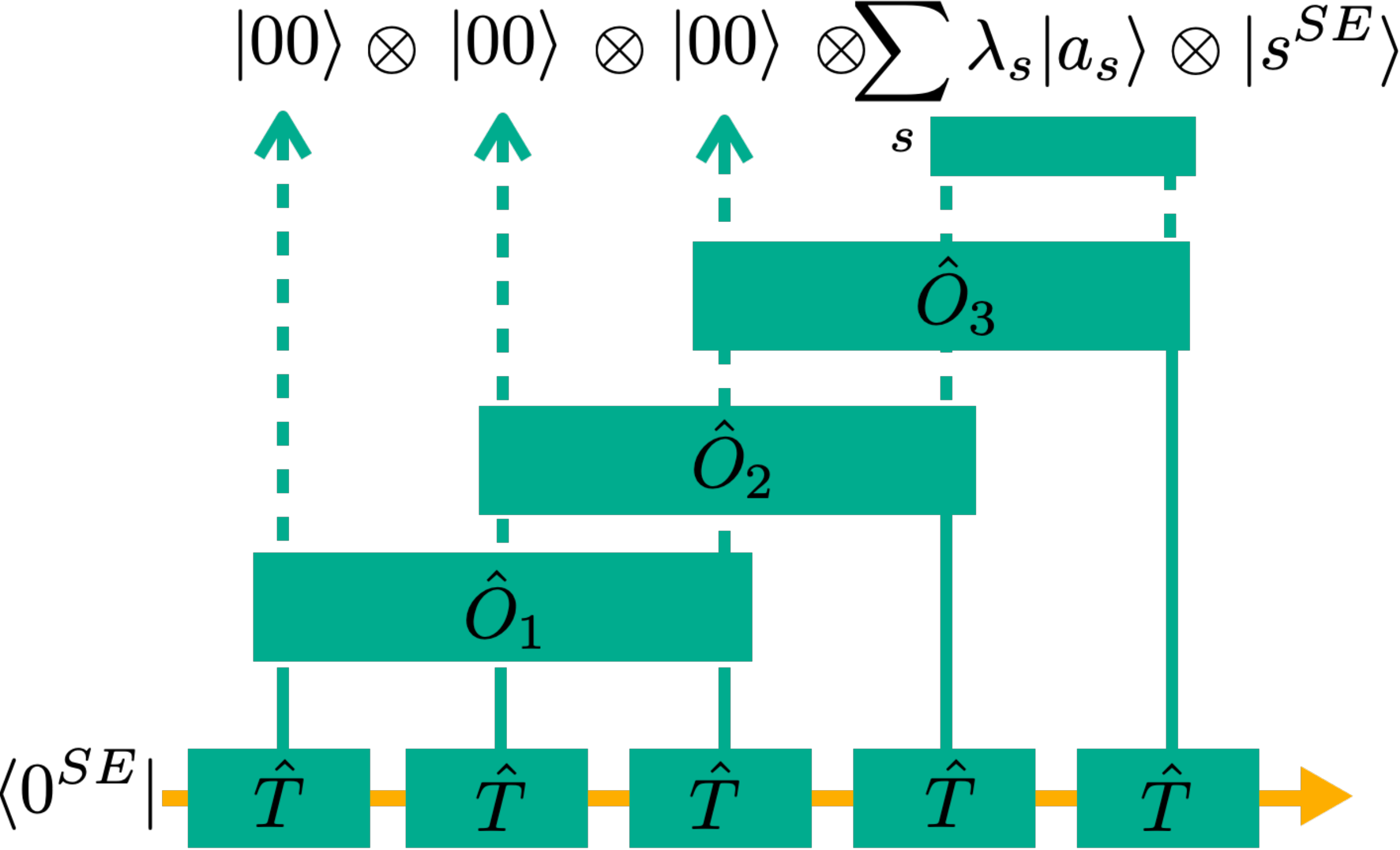}
\caption{Algorithm to reconstruct a $5$-step purified process tensor, where we have assumed that the environment size $D\leq d^4$. $\vert a_s\rangle$ denotes a set of orthogonal states for the last two sites, which can be fixed together with $\lambda_s$ by tomography of the reduced density operator on these two sites. The operator $\hat{O}_j$ is the $j$-th disentangling operator defined in Ref.~\cite{CramerLiu2010}.
}
\label{fig:fig2}
\end{figure}

PPT is only practically useful if it can be efficiently reconstructed based on experimental measurements. At first sight such a tomography algorithm seems unlikely due to the unknown initial state $\vert \psi^{SE}\rangle$ and the output environment state in Eq.(\ref{eq:ppt}). However, as will be shown, these unknown DOFs can be fixed by the invariance properties of PPT. 
Assuming that the unknown quantum process has a hidden OQE description in which the environment size is bounded by $D$, then we can apply a disentangling quantum circuit onto the (unknown) PPT and get 
\begin{align}\label{eq:tomo}
&\hat{O}_{f} \dots \hat{O}_2 \hat{O}_1 \ppt \vert \psi^{SE}\rangle =  \vert 0^{f:1}\rangle \otimes \sum_{s}\lambda_s \vert a_s\rangle \vert b_s \rangle,
\end{align}
with $R = \lceil\log_d(D)\rceil + 1$ and $f = N-R+1$. $\hat{O}_j$ is the $j$-th disentangling gate defined in Ref.~\cite{CramerLiu2010}, which requires tomography of the reduced density operator on the $j$ to $j+R-1$-th sites (Site $n$ means the pair of indices $(o_n, i_n)$). $\vert a_s\rangle$ is a set of unknown orthogonal states for the last $N-R$ sites and $\lambda_s$ are the Schmidt numbers. $\vert a_s\rangle$ and $\lambda_s$ can be fixed by tomography of the reduced density operator on the last $N-R$ sites (for which $\vert a_s\rangle$ is the eigenvector corresponding to the eigenvalue $\lambda_s^2$). $\vert b_s\rangle$ is a set of unknown orthogonal states for the environment. Since we can arbitrarily change the basis for the output environment state without any observable effects, we can simply fix this freedom by selecting $\vert b_s\rangle$ as the computational basis, that is, $\vert s^E\rangle$. Then the quantum state on the right hand side of Eq.(\ref{eq:tomo}) would be fixed. We can thus obtain the MPS representation of the PPT by reversing those gates, and then convert it into PPT (See Supplementary for details of this conversion). This tomography algorithm is demonstrated in Fig.~\ref{fig:fig2}. The resulting PPT completely characterizes the unknown quantum process since the hidden OQE is simply related to the PPT (Eq.(\ref{eq:Tmat})). For time-independent quantum process, one can further use the obtained PPT to predict any future process.

\section{Memory complexity of quantum process}
The OQE description hidden for a generic quantum process closely resembles the $\epsilon$-machine for a generic classical stochastic process. A classical stochastic process is fully specified by the conditional probability $P(X_N|X_{N-1:1})$ for any $N$, where $X_N$ is a random variable at time step $N$ and $X_{N-1:1}$ denotes all the histories, that is, the sequence of random variables from $X_1$ to $X_{N-1}$. The information needed to store $P(X_N|X_{N-1:1})$ grows exponentially with $N$, which is a problem that $\epsilon$-machine aims to solve. Instead of storing all $P(X_N|X_{N-1:1})$, $\epsilon$-machine divides $X_{N-1:1}$ into disjoint classes. We denote the set of all those distinct classes as $\epsilon(X_{N-1:1})$. Each element $\epsilon(x_{N-1:1}) \in \epsilon(X_{N-1:1})$ is itself a class of histories grouped together by the equivalence relation $P(X_N|x_{N-1:1}') = P(X_N|x_{N-1:1})$, where $x_{N-1:1}$ and $x_{N-1:1}'$ denote specific instances of the history. Therefore, to predict an output at time step $N$, one only needs to specify the current history $\epsilon(x_{N-1:1})$ together with the \textit{causal transition tensor} $P(\epsilon(X_{N:1}) = \epsilon(x_{N:1}), X_N=x_N| \epsilon(X_{N-1:1}) = \epsilon(x_{N-1:1}) )$, that is, the probability to obtain $x_N$ at time step $N$ and at the same time $\epsilon(x_{N-1:1})$ transits into $\epsilon(x_{N:1})$, given $\epsilon(x_{N-1:1})$. The future is thus fully determined by $\epsilon(x_{N-1:1})$, and for this reason the classes $\epsilon(x_{N-1:1})$ are referred to as the causal states or memory states~\cite{Shalizi2001}. A central observation is that although the number of histories grows exponentially with $N$, the number of memory states may not. 
Now for a generic quantum process, the environment plays the role of the memory space spanned by all the memory states. Given the environment state $\vert \alpha\rangle$ at time step $N-1$, the isometry $\hat{T}$ specifies the transition amplitude that the next state at time step $N$ is $\vert o_N, i_N\rangle$, and that the environment state becomes $\vert \beta\rangle$. Therefore $\hat{T}$ plays a similar role to the causal transition tensor.

The memory states for a classical stochastic process undergoes Markovian dynamics, since it is fully determined by the previous memory states. In fact the transition matrix for the memory states can be straightforwardly obtained by marginalizing the causal transition tensor over the `physical' index $x_N$.
Therefore the memory states will converge to the stationary distribution regardless of the initial state (as long as the stochastic process is ergodic), and then the future process will bear no memory of the initial state any more. Now from the PPT of a quantum process, the future process at time step $N$ and after is fully determined by the environment state at time step $N-1$, defined as
\begin{align}\label{eq:rho}
\rho_{N-1} = \overleftarrow{E}_{N-1} \dots \overleftarrow{E}_2    \overleftarrow{E}_1(\rho_0),
\end{align}
with $\rho_0 = \vert \psi^{SE}\rangle \langle \psi^{SE}\vert$ the system-environment initial state. $\transfer_n$ is the $n$-th transfer matrix of PPT, defined as $\transfer_n = \sum_{o_n, i_n} (B^{o_n, i_n})^{\ast} \otimes B^{o_n, i_n}$. The action of $\transfer_n$ on a state from the left is 
\begin{align}
\overleftarrow{E}(\rho) &= \sum_{o, i} \left(B^{o, i}\right)^{\dagger} \rho B^{o,i},
\end{align}
where the subscript $n$ has been omitted. Each $\rho_n$ is a $D\times D$ density operator of the environment. If the quantum process is time-independent, then Eq.(\ref{eq:rho}) can be simply written as $\rho_{N-1} = \overleftarrow{E}^{N-1}(\rho_0)$. In this case $\rho_N$ will converge to the left dominate eigenvector of $\transfer$ for large enough $N$, denoted as $\rhoss$, and the PPT after time step $N$ can then be described by an infinite MPS (iMPS) for which the initial state no longer matters~\cite{Schollwock2011,Orus2014}. This is in exact correspondence with the classical case. Formally, we can define a stationary quantum process (SQP) as follows: there exists a large enough integer $N_0 > 0$, such that the quantum process after it satisfies
\begin{align}\label{eq:stationarydef}
\pt_{N+L:M+L} = \pt_{N:M} \quad \forall N > M \geq N_0, \forall L >0,
\end{align}
where $\pt_{N:M}$ means the partial process tensor by tracing out the time steps from $1$ to $M-1$ from the process tensor $\pt_{N:1}$. Since $\pt_{N:M}$ is fully determined by $\rho_{M-1}$, Eq.(\ref{eq:stationarydef}) will be satisfied once $\rho_{M-1}$ has converged to $\rhoss$.

\begin{figure}
\includegraphics[width=\columnwidth]{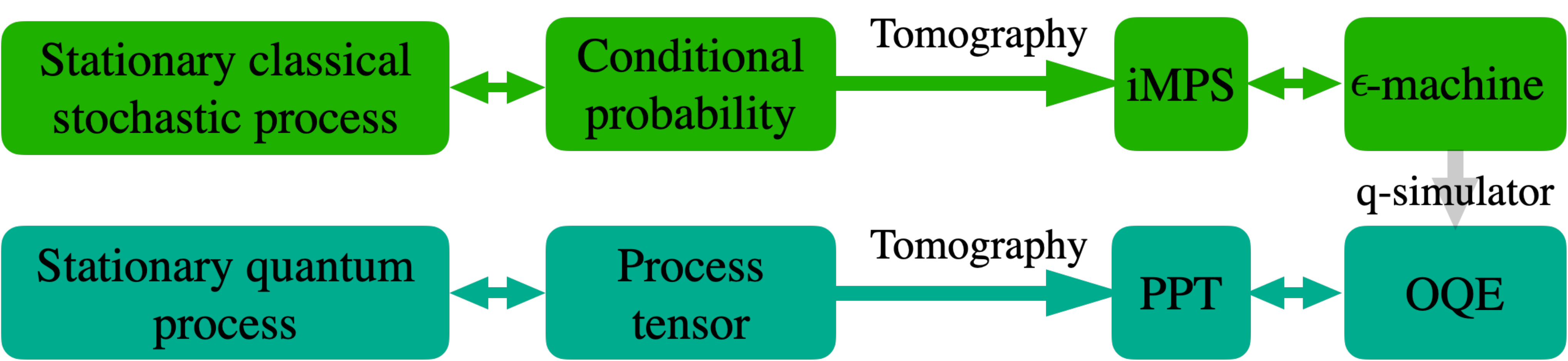}
\caption{The intimate connection between quantum processes and classical stochastic processes. A classical stochastic process is described by the condition probabilities $P(X_N|X_{N-1:1})$ or a hidden $\epsilon$-machine which can not be directly measured experimentally. The iMPS is an intermediate representation connects these two in that, it can be reconstructed from fixed-ranged conditional probabilities and it is equivalent to an $\epsilon$-machine. Similarly, a quantum process is fully described by the process tensor $\pt_{N-1:1}$, or a hidden OQE. PPT connects these two in the same way as the iMPS in the classical case. Finally, the q-simulator provides a way to simulate a stationary classical stochastic process on a quantum computer.
}
\label{fig:fig3}
\end{figure}

The $\epsilon$-machine for a stationary classical stochastic process (SCSP) is fully specified by the equivalence relation $\epsilon$ and the causal transition tensor. And we have shown that a SQP is fully specified by the choice of a basis for the environment and the isometry $\hat{T}$ (therefore it is fully specified by PPT). To this end, we also note that it has been shown that $\epsilon$-machine could also be described by an OQE (referred to as the q-simulator in the original literature)~\cite{BinderGu2018}, and that it has an equivalent iMPS representation~\cite{YangGu2018}. Importantly, the iMPS representation of a SCSP leads to a natural definition of the quantum memory complexity of a SCSP in terms of the quantum Renyi entropy of the iMPS~\cite{YangGu2018}, which is equivalent to the entanglement entropy of the stationary state in the memory space, therefore it can be physically interpreted as the resources required to simulate a SCSP on a quantum computer. Interestingly, the quantum memory complexity for a SCSP can be significantly smaller than the corresponding classical memory complexity, defined as the Renyi entropy of the stationary distribution of the classical memory states~\cite{ElliottGu2018,ElliottGu2020}. The intimate connection between SCSPs and SQPs is shown in Fig.~\ref{fig:fig3}.
Given these connections, we define the memory complexity of a SQP as
\begin{align}\label{eq:renyi} 
\mem^{\alpha} = \frac{1}{1-\alpha} \log_2\left(\trace\left(\rhoss^{\alpha}\right)\right),
\end{align}
which can be efficiently computed given the PPT. We further define the Schmidt rank of PPT as the memory size, which is the environment size in the minimal OQE description that produces the same quantum process. We have the following Theorem for $\mem^{\alpha}$ of generic quantum processes:

\textbf{Theorem 1.} Assuming the dominate eigenstate of $\transfer$ is non-degenerate, then $\mem^{\alpha} = \log_2 (D)$ if the $\vert \psi^{SE}\rangle$ is separable, and $ \mem^{\alpha} = \mathcal{C}_0^{\alpha} + \log_2 (D)$ if $\vert \psi^{SE}\rangle$ is entangled and the entropy of the reduced density operator $\rho^S$ is  $\mathcal{C}_0^{\alpha}$. Here $D$ means the size of the environment (not the effective one). Since the system here can also be any other DOFs that is mixed with the environment, the second part of the theorem covers the situation where the environment is initially in a mixed state with entropy $\mathcal{C}_0^{\alpha}$. 

\textit{Proof.} For separable $\vert \psi^{SE}\rangle$, it suffices to show that $\rhoss$ is the maximally mixed state, namely $\rhoss = \hat{I}^E/D$. For entangled $\vert \psi^{SE}\rangle$ as $\vert \psi^{SE}\rangle = \sum_s \lambda_s \vert x_s \rangle \vert y_s\rangle $, where $\vert x_s\rangle$ and $\vert y_s\rangle$ are sets of orthogonal basis for the system and environment respectively, and $\lambda_s$ are the Schmidt numbers, it suffices to show that $\rhoss = \sum_s \lambda_s^2 \vert x_s\rangle\langle x_s\vert \otimes \hat{I}^E / D $. More details of the proof can be found in Supplementary. Interestingly, based on Theorem 1, one could use $\mem^{\alpha}$ to detect the memory size since the former is experimentally accessible~\cite{DaleyZoller2012,IslamGreiner2015}.

In summary, we have defined the purified process tensor for generic quantum processes, which contains all the information of a generic quantum process and is naturally a sequentially generated multi-qubit state. An efficient tomography algorithm for PPT is shown. Drawing the intimate connection between PPT and the $\epsilon$-machine for classical stochastic processes, we define the stationary quantum process as well as the memory complexity of it, which can be interpreted as the amount of resources required to simulate this quantum process on a quantum computer. The exact values of the memory complexities for generic quantum processes are given. 

\begin{acknowledgments}
C. G. would like to thank Chengran Yang for helpful discussions. C. G. acknowledges support from National Natural Science Foundation of China under Grant No. 11805279.
\end{acknowledgments}

\bibliographystyle{apsrev4-1}
\bibliography{refs}

\appendix

\section{Conversion between PPT and its MPS Representation}
It has been shown that there is a one-to-one correspondence between a sequentially generated multi-qubit state (SGMS) with environment size $D$, and an MPS with bond dimension $D$~\cite{SchonWolf2005}. Now assuming that we apply an isometry $\hat{S}$ on the environment, then the initial state and the isometry $\hat{T}$ will transform as
\begin{align}
\vert (\psi^{SE})'\rangle &= \hat{S} \vert \psi^{SE}\rangle; \\
\hat{T}' &= \hat{S} \hat{T} \hat{S}^{\dagger},
\end{align}
as a result, under such an isometry the PPT will become
\begin{align}
\hat{T}_N' \dots \hat{T}_2' \hat{T}_1' \vert (\psi^{SE})^{\prime} \rangle &= \hat{S}  \hat{T}_N \hat{S}^{\dagger} \dots \hat{S}\hat{T}_2 \hat{S}^{\dagger} \hat{S} \hat{T}_1 \hat{S}^{\dagger} \hat{S} \vert \psi^{SE} \rangle \nonumber \\ 
&= \hat{S}\hat{T}_N \dots \hat{T}_2 \hat{T}_1 \vert \psi^{SE} \rangle.
\end{align}
The last isometry $\hat{S}$ on PPT will not have any observable effects since this index corresponds to the output environment, and is going to be contracted when computing observables. Therefore the PPT is invariant under any isometry of the environment, or that there exists infinite many hidden OQE which will result in equivalent PPTs. 

Given the PPT, the corresponding MPS can be read as 
\begin{align}
\sum_{\alpha_{N-1:1}, o_{N:1}, i_{N:1}} B^{o_1, i_1}_{\alpha_1} \cdots B^{o_N, i_N}_{\alpha_{N-1}, \alpha_N} \vert o_1, i_1, \dots, o_N, i_N\rangle,
\end{align}
with the site matrix $B^{o_n, i_n}$ defined as
\begin{align}\label{eq:sitem}
B^{o_n, i_n}_{\alpha_{n-1}, \alpha_n} = \frac{1}{\sqrt{d}} U^{o_n, \alpha_n}_{i_n, \alpha_{n-1}}, \quad \forall n > 1,
\end{align}
and $B^{o_1, i_1}_{\alpha_1}$ is a row vector given by 
\begin{align}
B^{o_1, i_1}_{\alpha_1} = \sum_{\alpha_0} \frac{1}{\sqrt{d}} U^{o_1, \alpha_1}_{i_1, \alpha_{0}} \nu_{\alpha_0}.
\end{align}
Here $\nu_{\alpha_0} = \langle \alpha_0 \vert \psi^{SE}\rangle$ is the element of the coefficient vector corresponding to the initial state $\vert \psi^{SE}\rangle$. Now we show that $\hat{T}$ in Eq.(\ref{eq:Tmat}) is an isometry, which can be proven by
\begin{align}\label{eq:rightcanonical}
&\langle 00\vert \langle \alpha' \vert \hat{T}^{\dagger} \hat{T}\vert \alpha \rangle \vert 00\rangle \nonumber \\
= &\frac{1}{d} \sum_{\beta', k', j'} (U^{j', \alpha'}_{k', \beta'})^{\ast}\langle \beta', k', j'\vert \sum_{\beta, k, j} U^{j, \alpha}_{k, \beta} \vert \beta, k, j\rangle \nonumber \\ 
=& \frac{1}{d} \sum_{j, j'} \delta_{j, j'} \left(\sum_{\beta, k} (U^{j', \alpha'}_{k, \beta})^{\ast} U^{j, \alpha}_{k, \beta}\right) \nonumber \\
=& \frac{1}{d} \sum_{j, j'} \delta_{j, j'}^2 \delta_{\alpha, \alpha'} = \delta_{\alpha, \alpha'},
\end{align}
where $\delta$ is the Kronecker delta function. 
Similarly, the site matrix $B^{o_n, i_n}$ is right-canonical~\cite{Schollwock2011}, that is
\begin{align}
\sum_{o_n, i_n}(B^{o_n, i_n, \alpha_n}_{\alpha_{n-1}', \alpha_n})^{\ast} B^{o_n, i_n}_{\alpha_{n-1}, \alpha_n} = \delta_{\alpha_{n-1}', \alpha_{n-1}}.
\end{align}
Moreover, a similar procedure can prove that $B^{o_n, i_n}$ is also left-canonical except for $n=1$. This is due to the fact that the two indices $\alpha_n$ and $\alpha_{n-1}$ are of the same priority as can be seen from Eq.(\ref{eq:sitem}). However on the first site the initial condition breaks the left-canonicality of $B$.

Now we show that given an MPS obtained through tomography, denoted as $\ppt_{N:1}^t \vert 0^{SE}\rangle$, one can directly get the PPT, thus get $\evo_n$ for each time step.
First we can prepare the MPS in the right canonical form as
\begin{align}
&\ppt_{N:1}^t \vert 0^{SE}\rangle \nonumber \\ 
=& \sum_{\alpha_{N-1:1}, o_{N:1}, i_{N:1}} \tilde{B}^{o_1, i_1}_{\alpha_1} \cdots \tilde{B}^{o_N, i_N}_{\alpha_{N-1}, \alpha_N} \vert o_1, i_1, \dots, o_N, i_N\rangle,
\end{align}
then each $\tilde{B}^{o_n, i_n}_{\alpha_{n-1}, \alpha_n}$ is naturally an isometry from $\mathcal{H}^E$ to $\mathcal{H}^E \otimes \mathcal{H}^S \otimes \mathcal{H}^S$, where $\mathcal{H}^{E, S}$ means the Hilbert spaces of the environment and the system respectively. Then one can reshape $\tilde{B}$ to obtain
\begin{align}\label{eq:mpstoppt}
\tilde{M}^{o_n, \alpha_n}_{i_n, \alpha_{n-1}} =  \tilde{B}^{o_n, i_n}_{\alpha_{n-1}, \alpha_n}.
\end{align}
However, it is not guaranteed that $\tilde{M} = \tilde{U} / \sqrt{d} $ with $\tilde{U}$ some unitary operation on the space $\mathcal{H}^E \otimes \mathcal{H}^S$, by only supplying a right canonical MPS. This can only be guaranteed by the physics behind, since the underlying MPS originates from a hidden OQE. This can also be seen as follows. The site matrices for the right canonical MPS is unique up to a unitary transformation on the auxiliary (environment) degrees of freedom, thus $\tilde{M}$ will be different from the hidden unitary $\frac{1}{\sqrt{d}} \tilde{U}$ by at most a unitary transformation on the environment, which will still be a unitary operation on the system and environment. 

If the tomography is not exact, then the matrix $\tilde{M}$ computed with Eq.(\ref{eq:mpstoppt}) may not be strictly unitary. In this case one could also devise a variational ansatz for the PPT, which explicitly preserves the unitary property of $\evo_n$, as 
\begin{align}\label{eq:ppt_ansatz}
\ppt^a_{N:1} \vert 0^{SE}\rangle = \hat{O}^f \hat{T}_N \dots \hat{T}_2 \hat{T}_1 \vert 0^{SE}\rangle,
\end{align}
where $\hat{T}_n$ is related to $\evo_n$ (for which the corresponding tensor $U$ is parametric) by Eq.(\ref{eq:Tmat}) and we have fixed the initial state to be $\vert 0^{SE}\rangle$ without loss of generality. $\hat{O}^f$ is also a parametric unitary matrix with size $D\times D$ acting on the (effective) environment. Then one can optimize those $(N+1)$ parametric unitary matrices by minimizing a loss function, such as,
\begin{align}\label{eq:loss}
\loss(\evo_1, \dots, \evo_N, \hat{O^f}) = | \ppt^t \vert 0^{SE}\rangle - \ppt^a_{N:1} \vert 0^{SE}\rangle |^2,
\end{align}
where $|\cdot|$ means square of the Euclidean norm. 
For time-independent quantum process, one simply uses the same parametric $\evo$ for all time steps $1$ to $N$ in Eq.(\ref{eq:ppt_ansatz}), and once it is found, any future process can be easily predicted. This algorithm, however, is no longer deterministic and could well be trapped in some local minima. In practice one could use a unitary matrix which is closest to $\tilde{M}$ in Eq.(\ref{eq:mpstoppt}) as the initial parameters for Eq.(\ref{eq:loss}) to accelerate convergence as well as increasing the success probability.

\section{Computing observables with PPT}

\begin{figure}
\includegraphics[width=\columnwidth]{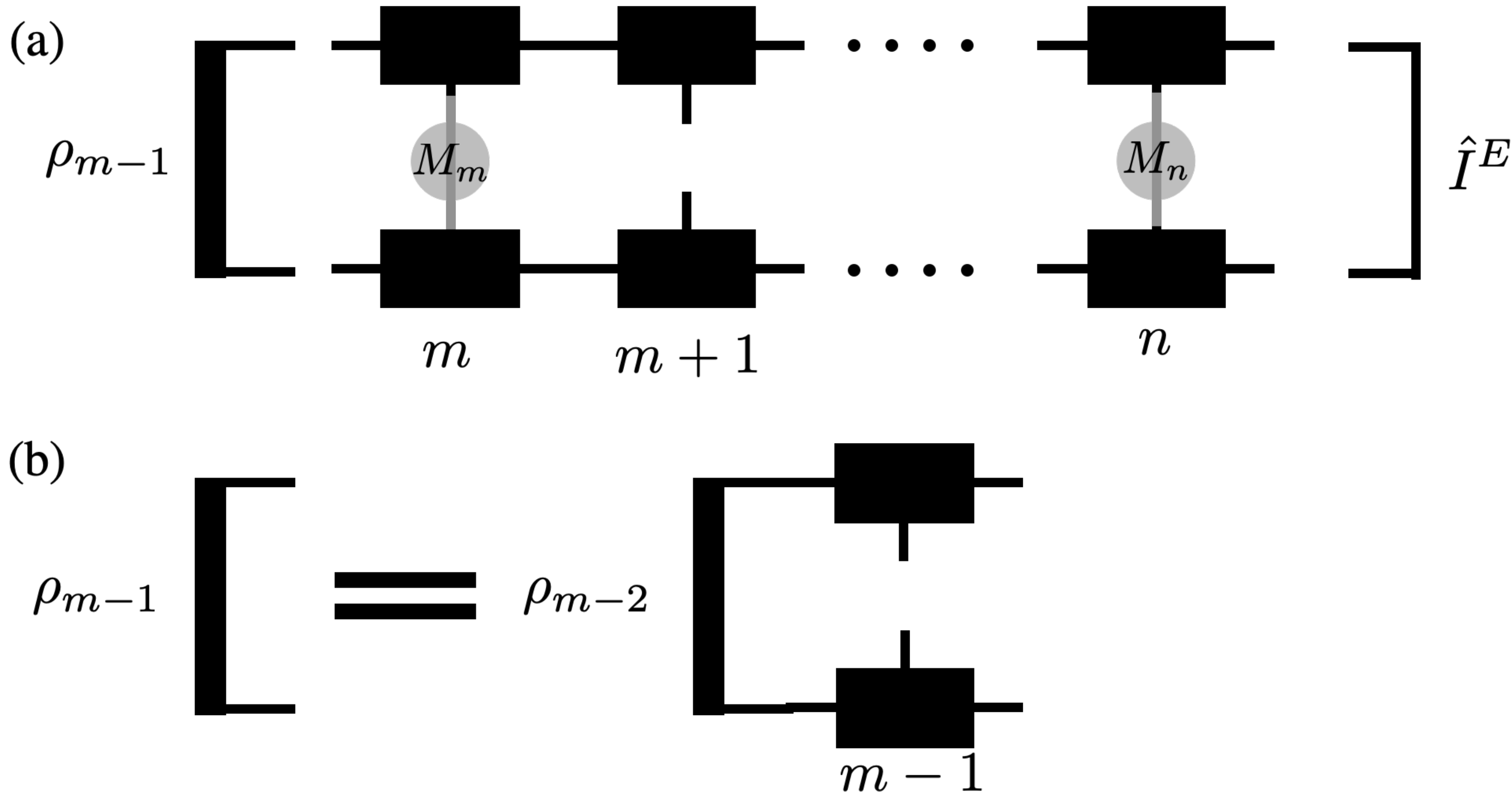}
\caption{(a) Computing a two-time correlation specified by two operators $M_m$ and $M_n$ at time steps $m$ and $n$. (b) Updating of the environment state from thee left.
}
\label{fig:figS1}
\end{figure}

Given the PPT, one could compute any multi-time correlations using the standard approaches to compute observables with MPS representation of pure states, which is shown in Fig.~\ref{fig:figS1}. Taking a two-time correlation with two operators $M_m$, $M_n$ acting on the $m$ and $n$-th time step (assuming $m < n$) as an example ($M_n$ is a $d^2\times d^2$ matrix which may be decomposed as a preparation on the input index $i_o$ followed by a measurement on the output index $o_n$, one can refer to Ref.~\cite{PollockModi2018a} for the physical correspondence of such an operator). We will also use a single index $\sigma_n$ to denote the index pair $(o_n, i_n)$ for briefness. Since the site matrix $B^{\sigma}$ is right canonical, we do not need to care about the site matrices with time steps larger than $n$. Mathematically, the expectation value can be evaluated with 
\begin{align}\label{eq:expec}
&\langle \psi^{SE}\vert \pptd M_m M_n \ppt \vert \psi^{SE}\rangle \nonumber \\ 
= &\sum_{\alpha_{n-1:m-1}, \alpha_{n-1:m-1}', \sigma_{m}, \sigma_m', \sigma_n,\sigma_n', \alpha_n } \rho_{\alpha_{m-1}, \alpha_{m-1}'} \nonumber \\ 
&\times \left((B^{\sigma_m'}_{\alpha_{m-1}', \alpha_m'})^{\ast} M_m^{\sigma_m', \sigma_m} B^{\sigma_m}_{\alpha_{m-1}, \alpha_m} \right) \nonumber \\ 
&\times E^{\alpha_{m}', \alpha_{m+1}'}_{\alpha_{m}, \alpha_{m+1}}  \times \cdots E^{\alpha_{n-2}', \alpha_{n-1}'}_{\alpha_{n-2}, \alpha_{n-1}} \nonumber \\ 
&\times \left((B^{\sigma_n'}_{\alpha_{n-1}', \alpha_n})^{\ast} M_n^{\sigma_n', \sigma_n} B^{\sigma_n}_{\alpha_{n-1}, \alpha_n} \right) ,
\end{align}
Here $E^{\alpha_{l-1}', \alpha_{l}'}_{\alpha_{l-1}, \alpha_l}$ is the transfer matrix at the $l$-th time step, as defined in the main text, whose explicit form is
\begin{align}
E^{\alpha_{l-1}', \alpha_{l}'}_{\alpha_{l-1}, \alpha_l} = \sum_{\sigma_l} (B^{\sigma_{l}}_{\alpha_{l-1}', \alpha_{l}'})^{\ast} B^{\sigma_{l}}_{\alpha_{l-1}, \alpha_{l}}.
\end{align}
$\rho_{\alpha_{m-1}', \alpha_{m-1}}$ denotes the environment state after time step $m-1$ (which is the matrix form of the state $\rho_{m-1}$ defined in the main text), defined as
\begin{align}
\rho_{\alpha_{m-1}', \alpha_{m-1}} = \sum_{\alpha_{m-2}, \alpha_{m-2}'} E^{\alpha_{m-2}', \alpha_{m-1}'}_{\alpha_{m-2}, \alpha_{m-1}}  \rho_{\alpha_{m-2}', \alpha_{m-2}},
\end{align}
with $\rho_{\alpha_0', \alpha_0} = \nu_{\alpha_0}^{\ast}\nu_{\alpha_0}$. The environment contribution from the right is simply identity due to the right canonical form of the $B$ matrices (also because the starting point for the environment state from right is the maximally mixed state since we will not do anything to the environment). As a result, we can see that for any observables which only contains operations on time steps starting from $m$, all the effects from the $1$ to $m-1$-th time steps are included in the environment state $\rho_{m-1}$, or that the future process at time step $m$ and after is fully determined by $\rho_{m-1}$.


\section{Infinite MPS for stationary quantum process}

\begin{figure}
\includegraphics[width=\columnwidth]{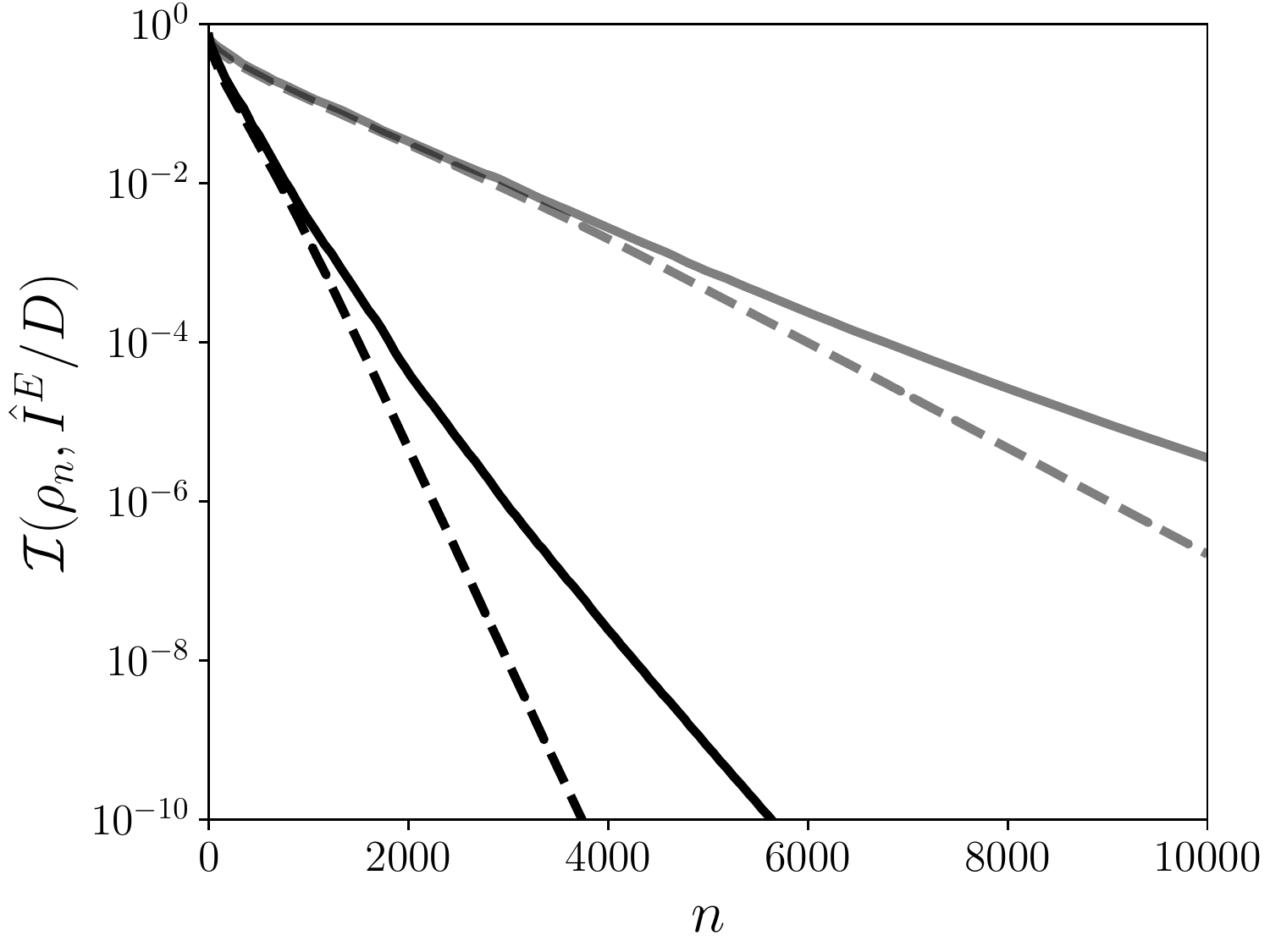}
\caption{The infidelity, defined as $\mathcal{I}(A, B) = 1 - \mathcal{F}(A, B)$ for two density matrices $A$ and $B$ with $\mathcal{F}$ to be the quantum fidelity, between the environment state at the $n$-th time step, namely $\rho_n $ and the maximally mixed state $\hat{I}/D$, as a function of the time step $n$. The unitary operator is generated by $\evo = \hat{I} + \eta \hat{H}$ with $\hat{H}$ a Hermitian matrix randomly generated using normal distribution. We have used a fixed random $\hat{H}$ for all time steps in the solid lines and a randomly generated $\hat{H}$ for each time step $n$ in the dashed lines. For the black solid and dashed lines we have used $\eta=0.01$, while for the gray solid and dashed lines we have used $\eta=0.005$.
}
\label{fig:figS2}
\end{figure}

In case $\evo$ is time-independent, $\rho_m$ will converges to the left dominate eigenvector of $\transfer$ with the largest eigenvalue, which is usually assumed to be non-degenerate~\cite{Schollwock2011} (we will also explicitly consider a case where the largest eigenvalue is degenerate later). Now we proceed to prove Theorem 1 in the main text.

First we prove that any left eigenvector of $\transfer$ has eigenvalue smaller or equal to $1$ (then it will be straightforward to show that it is also true for the right eigenvectors since there is no priority for the indices $\alpha_{m-1}$ and $\alpha_m$ from Eq.(\ref{eq:sitem})). For any basis $\vert a\rangle \langle b \vert$ of the density operator of the environment, we have
\begin{align}\label{eq:E1}
|\overleftarrow{E}(\vert a\rangle\langle b\vert)|^2 = \sum_{\sigma, \sigma'} (B^{\sigma}_{a, \alpha'})^{\ast} B^{\sigma}_{b, \alpha} B^{\sigma'}_{a, \alpha'} (B^{\sigma'}_{b, \alpha})^{\ast},
\end{align}
where $|\cdot|^2$ means the square of the Euclidean norm.
Now we define two matrices 
\begin{align}
X^{\sigma, \sigma'}_a &= \sum_{\alpha'} B^{\sigma}_{a, \alpha'} (B^{\sigma'}_{a, \alpha'})^{\ast} ; \\
Y^{\sigma, \sigma'}_b &= \sum_{\alpha} B^{\sigma}_{b, \alpha} (B^{\sigma'}_{b, \alpha})^{\ast},
\end{align}
which are semi-positive and Hermitian matrices by definition. We can also see that $\trace(X) = \sum_{\alpha', \sigma} B^{\sigma}_{a, \alpha'} (B^{\sigma}_{a, \alpha'})^{\ast} = 1 $ due to the right-canonicality of $B$, and the same for $Y$.
Then Eq.(\ref{eq:E1}) can be written as
\begin{align}\label{eq:proof1}
|\overleftarrow{E}(\vert a\rangle\langle b\vert)|^2 &= \trace(X^{\dagger} Y) \leq \sqrt{\trace(X^{\dagger} X)}\sqrt{\trace(Y^{\dagger} Y)} \nonumber \\ 
&= \sqrt{\trace(X^2)} \sqrt{\trace(Y^2)} \nonumber \\ 
&\leq \sqrt{\trace^2(X)} \sqrt{\trace^2(Y)} \nonumber \\ 
&= \trace(X) \trace(Y) = 1 ,
\end{align}
where the second step in the first line of Eq.(\ref{eq:proof1}) follows from the Cauchy-Schwarz inequality and the inequality in the second line is due to the semi-positivity of $X$ and $Y$. Equality holds only if $a = b$. Thus for any state $\rho = \sum_{a, b}\rho_{a, b}\vert a\rangle\langle b\vert$, we have
\begin{align}
|\overleftarrow{E}(\rho)|^2 = |\sum_{a, b} \rho_{a, b} \overleftarrow{E}(\vert a\rangle\langle b\vert)|^2 \leq |\rho|^2 |\overleftarrow{E}(\vert a\rangle\langle b\vert)|^2  = |\rho|^2, 
\end{align}
therefore any left eigenvector of $\transfer$ has an eigenvalue that is not greater than $1$.

Second we show that the maximally mixed state $\hat{I}^E / D$ is both a left and right eigenvector of $\transfer$ with eigenvalue $1$. This immediately follows since $B^{\sigma}$ is both left and right-canonical (We do not care about the first site, $B^{\sigma_1}$, for infinite case). Thus the first part of Theorem 1 is proved. The convergence of the environment state to the maximally mixed state in this case is also numerically demonstrated in Fig.~\ref{fig:figS2}.

Now we proceed to prove the second part of Theorem 1 for entangled system-environment initial state. In this case the site matrix corresponding to the effectively enlarged environment is 
\begin{align}
(B^{\sigma})^{\prime} = B^{\sigma} \otimes I^E,
\end{align}
which is certainly degenerate. However, in this case we will show in the following that the quantum process can still be stationary. First we note that the traceless matrices span a linear subspace which is orthogonal to the maximally entangled state and that $\transfer$ matrix only maps traceless matrix to traceless matrix due to trace preservation, then since the largest eigenvalue is assumed to be non-degenerate, all the traceless matrices have eigenvalues that are strictly less than $1$. Now in case the initial state $\vert \psi^{SE}\rangle$ is entangled, we assume that it can be decomposed as
\begin{align}\label{eq:entangled_state}
\vert \psi^{SE}\rangle = \sum_{s}\lambda_s \vert x_s\rangle \vert y_s\rangle
\end{align}
without loss of generality. Here $\vert x_s\rangle$ and $\vert y_s\rangle$ are sets of orthogonal basis for the system and environment respectively, and $\lambda_s$ are the Schmidt numbers. Therefore $\rho_0$ can be written as
\begin{align}
\rho_0 =  \sum_{s, s'} \lambda_s \lambda_{s'} \vert x_s\rangle\langle x_{s'}\vert \otimes \vert y_s\rangle  \langle y_{s'}\vert.
\end{align}
Now we have
\begin{align}
\overleftarrow{E}^n(\rho_0) = \sum_{s, s'} \lambda_s \lambda_{s'} \vert x_s\rangle\langle x_{s'}\vert \otimes \overleftarrow{E}^n(\vert y_s\rangle  \langle y_{s'}\vert).
\end{align}
However any $s\neq s'$, the action $\overleftarrow{E}^n$ on it will vanish for large enough $n$, therefore we are only left with the diagonal terms. Moreover, $\overleftarrow{E}^n$ acting on pure states $\vert s\rangle \langle s\vert $ will converge to the maximumly mixed state. As a result, we have
\begin{align}
\overleftarrow{E}^{n \rightarrow \infty}(\rho_0) = \sum_s \lambda_s^2 \vert x_s\rangle\langle x_{s}\vert \otimes \hat{I}^E / D,
\end{align}
for which the entropy has the desired form as in Theorem 1. Therefore, even though the dominate eigenspace is degenerate, the initial state will automatically picks a particular one from it as the stationary state.

\section{Reconstruction of the entangled system-environment initial state}

\begin{figure}
\includegraphics[width=\columnwidth]{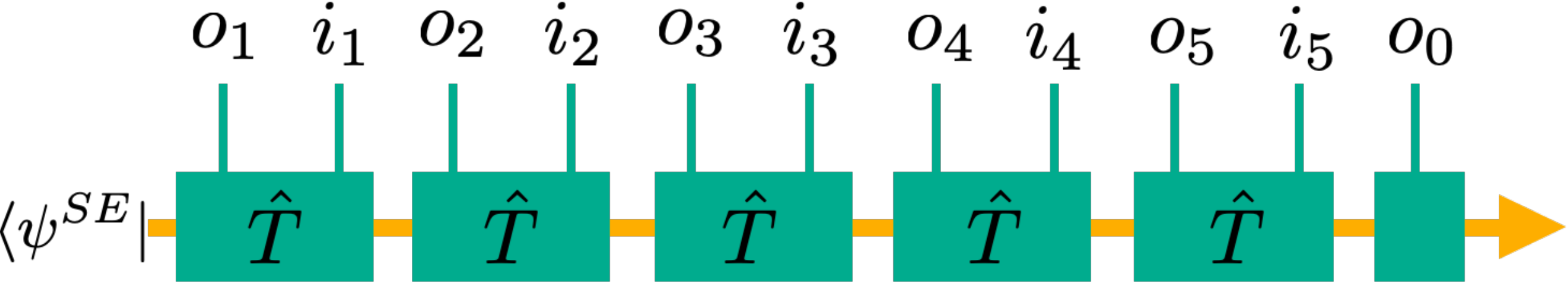}
\caption{Reconstruction of the PPT in presence of the initial system-environment entanglement, where the system index corresponding to the initial state $o_0$ can be further split off from the `effective' environment, since one can directly perform measurements on the initial state of the system.
}
\label{fig:figS3}
\end{figure}

The convention of our PPT treats the index $o_0$ corresponding to the initial state of the system on the same footing as the environment, which means that when computing observables, no operations (measurements) are made on the initial state corresponding to $o_0$. As a result, the information about future process conditioned on the measurement on the initial state is lost (or ignored by averaging out). The reason for this choice is simply for easy visualization of the connection between the PPT and a SGMS. To recover this information, we can simply perform tomography on the modified `PPT' as shown in Fig.~\ref{fig:figS3}. The tomography of the modified PPT can be done in the same way using the algorithm shown in the main text, except in the end one can further split off the system index $o_0$ from the effective environment. However, the isometry $\hat{T}$ obtained with this approach will actually be the enlarged one acting on the enlarged effective environment plus the system, that is, $\hat{T}\otimes \hat{I}^S$ (more concretely, it will be of size $d^2D\times d^2D$ instead of $dD\times dD$).

To directly recover $\hat{T}$ that acts on the true environment plus the system and at the same time with no loss of information on the initial state, one can proceed as follows.
Assuming the entangled initial state has the form in Eq.(\ref{eq:entangled_state}), we can first do a tomography for the system, which allows us to identify the states $\vert x_s\rangle$ together with the Schmidt numbers $\lambda_s$. Then we measure the system in the basis $\{\vert x_s\rangle\langle x_s\vert \}$, which will result in $\vert x_s\rangle$ with probability $\lambda_s^2$ and at the same time the system and environment are disentangled, leaving the environment in a pure state $\vert y_s\rangle$. Then we can proceed to reconstruct the PPT $\ppt_{N:1} \vert y_s\rangle$, with which we can further reconstruct the evolutionary operator $\evo_n$ by assuming the initial state of the environment is $\vert r(s)^E\rangle$ with $r(s)$ a predefined one-to-one index mapping. If we obtain another set of outcomes $\vert x_{s'}\rangle$, $\vert y_{s'}\rangle$, we fixed $\evo_n$ and use the following ansatz for the PPT
\begin{align}
\ppt^a \vert r(s')^E \rangle = \hat{O}^f \hat{T}_N \dots \hat{T}_2 \hat{T}_1 \vert  r(s')^E\rangle,
\end{align}
where $\hat{T}_n$ is the known isometry corresponding to $\evo_n$, while the initial state is parameterized in the space orthogonal to all the previous $\vert r(s)^E\rangle$, and $\hat{O}^f$ is still a parameterized unitary matrix. In this way we can fully determined the initial state as 
\begin{align}
\vert \psi^{SE}\rangle = \sum_s \lambda_s \vert x_s\rangle \vert r(s)^E\rangle.
\end{align}
Therefore all the information of the underlying quantum process are restored. 

\end{document}